\documentclass[twocolumn]{aastex631}

\newcommand\kms{km$\,$s$^{-1}$}
\newcommand\Msol{M$_{\odot}$}

\newcommand{\hi}{H\,{\sc i}}
\newcommand{\hii}{H\,{\sc ii}}

\graphicspath{{./}{figures/}}

\begin{document}

\title{Pavo: Stellar feedback in action in a low-mass dwarf galaxy}
\shorttitle{Pavo \hi}

\correspondingauthor{Michael G. Jones}
\email{jonesmg@arizona.edu}

\author[0000-0002-5434-4904]{Michael G. Jones}
\affiliation{Steward Observatory, University of Arizona, 933 North Cherry Avenue, Rm. N204, Tucson, AZ 85721-0065, USA}

\author[0000-0002-1515-995X]{Martin P. Rey}
\affiliation{Department of Physics, University of Bath, Claverton Down, Bath, BA2 7AY, United Kingdom}

\author[0000-0003-4102-380X]{David J. Sand}
\affiliation{Steward Observatory, University of Arizona, 933 North Cherry Avenue, Rm. N204, Tucson, AZ 85721-0065, USA}

\author[0000-0002-0956-7949]{Kristine Spekkens}
\affiliation{Department of Physics, Engineering Physics and Astronomy, Queen’s University, Kingston, ON K7L 3N6, Canada}

\author[0000-0001-9649-4815]{Bur\c{c}in Mutlu-Pakdil}
\affil{Department of Physics and Astronomy, Dartmouth College, Hanover, NH 03755, USA}

\author[0000-0002-9798-5111]{Elizabeth A. K. Adams}
\affiliation{ASTRON, the Netherlands Institute for Radio Astronomy, Oude Hoogeveensedijk 4,7991 PD Dwingeloo, The Netherlands}
\affiliation{Kapteyn Astronomical Institute, University of Groningen, PO Box 800, 9700 AV Groningen, The Netherlands}

\author[0000-0001-8354-7279]{Paul Bennet}
\affiliation{Space Telescope Science Institute, 3700 San Martin Drive, Baltimore, MD 21218, USA}

\author[0000-0002-1763-4128]{Denija Crnojevi\'{c}}
\affil{University of Tampa, 401 West Kennedy Boulevard, Tampa, FL 33606, USA}

\author[0000-0001-9775-9029]{Amandine~Doliva-Dolinsky}
\affil{Department of Physics, University of Surrey, Guildford GU2 7XH, UK}
\affiliation{Department of Physics and Astronomy, Dartmouth College, Hanover, NH 03755, USA}

\author[0000-0001-7618-8212]{Richard Donnerstein}
\affiliation{Steward Observatory, University of Arizona, 933 North Cherry Avenue, Rm. N204, Tucson, AZ 85721-0065, USA}

\author[0000-0001-8245-779X]{Catherine E. Fielder}
\affiliation{Steward Observatory, University of Arizona, 933 North Cherry Avenue, Rm. N204, Tucson, AZ 85721-0065, USA}

\author[0000-0003-1020-8684]{Julia Healy}
\affiliation{Jodrell Bank Centre for Astrophysics, School of Physics and Astronomy, University of Manchester, Oxford Road, Manchester M13 9PL, UK}

\author[0000-0001-5368-3632]{Laura C. Hunter}
\affiliation{Department of Physics and Astronomy, Dartmouth College, 6127 Wilder Laboratory, Hanover, NH 03755, USA}

\author[0000-0001-8855-3635]{Ananthan Karunakaran}
\affiliation{Department of Astronomy \& Astrophysics, University of Toronto, Toronto, ON M5S 3H4, Canada}
\affiliation{Dunlap Institute for Astronomy and Astrophysics, University of Toronto, Toronto ON, M5S 3H4, Canada}

\author[0000-0002-8217-5626]{Deepthi S. Prabhu}
\affiliation{Steward Observatory, University of Arizona, 933 North Cherry Avenue, Rm. N204, Tucson, AZ 85721-0065, USA}

\author[0000-0002-5177-727X]{Dennis Zaritsky}
\affiliation{Steward Observatory, University of Arizona, 933 North Cherry Avenue, Rm. N204, Tucson, AZ 85721-0065, USA}



\begin{abstract}
MeerKAT observations of the recently discovered, extremely low mass galaxy, Pavo, have revealed a neutral gas (\hi) reservoir that was undetected in archival \hi \ single dish data. We measure Pavo's \hi \ mass as $\log M_\mathrm{HI}/\mathrm{M_\odot} = 5.79 \pm 0.05$, making it the lowest mass \hi \ reservoir currently known in an isolated galaxy (with a robust distance measurement). Despite Pavo's extreme isolation, with no known neighbor within over 700~kpc, its \hi \ reservoir is highly disturbed. It does not show clear signs of rotation and its center of mass is offset from the stellar body center by 320~pc, while its peak is offset by 82~pc (both in projection). Despite this disturbed morphology, Pavo still appears to be consistent with the \hi \ size--mass relation, although it is not possible to accurately determine a suitable inclination correction. Such disturbed, offset and disorganized \hi \ reservoirs are predicted by simulations of low-mass, star-forming dwarfs in which supernova-driven outflows efficiently disrupt the interstellar medium after a star formation event. It is likely that we are witnessing Pavo in precisely this period, tens to a few hundred Myr after a star formation episode, when internal feedback has disrupted its gas reservoir.
\end{abstract}



\section{Introduction} \label{sec:intro}

The lowest mass star-forming galaxies in the nearby universe are of great interest as they provide a window to how star formation (SF) proceeds in an extremely metal-poor interstellar medium (ISM), analogous to conditions in the early universe \citep[e.g.,][]{Irwin+2007,Weisz+2012,Giovanelli+2013,Skillman+2013,McQuinn+2015a,McQuinn+2015b,McQuinn+2024,Sand+2015b,Jones+2023b,Jones+2024}. In addition, their gas kinematics has the potential to illuminate the inner structure of their dark matter (DM) halos, directly addressing the cusp/core controversy \citep[e.g.][]{deBlok+2010, Read+2016, Rey+2024}. However, identifying such galaxies has proven remarkably difficult and at present only a handful of star-forming galaxies with $M_\ast \lesssim 10^6$~\Msol \ are known. The canonical object in this class, Leo~P \citep{Giovanelli+2013}, was identified in the Arecibo Legacy Fast ALFA (Arecibo L-band Feed Array) survey, or ALFALFA \citep{Giovanelli+2005,Haynes+2018}, through the 21~cm \hi \ line emission of its neutral gas reservoir. 

Despite the expectation that these are the most numerous star-forming galaxies in the universe, only a few partial-analogs to Leo~P (e.g. Leo~T, Antlia~B, Leoncino) have been identified through resolved star searches or \hi \ surveys \citep{Irwin+2007,Sand+2015b,Hirschauer+2016}. This is likely the result of current ground-based resolved star searches being ineffective much beyond the sphere of influence of the Local Group (LG) \citep[e.g.][]{Mutlu-Pakdil+2021}, within which environmental processes appear to rapidly quench low-mass galaxies \citep[e.g.][]{Spekkens+2014,Putman+2021}. Similarly, existing \hi \ surveys \citep{Barnes+2001,Giovanelli+2005} lack either the sensitivity or sky coverage (or both) to probe significant volumes beyond the LG at this mass scale.
In both cases, these limitations are likely to be swept aside in the coming years. Optical/IR survey facilities such as the Rubin Observatory \citep{Ivezic+2019}, Euclid \citep{Euclid1}, and the Nancy Grace Roman Space Telescope \citep{Akeson+2019} will be capable of resolving stellar populations out to greater distances, while ongoing \hi \ surveys with the Five-hundred-meter Aperture Spherical Telescope (FAST)\footnote{We note that FAST did indeed recently identify a nearby, extremely low mass star-forming galaxy \citep{Xu+2025}. However, the distance to this object is highly uncertain and it may be several times more massive than its fiducial value.} and the Australian Square Kilometre Array Pathfinder (ASKAP) will exceed both the sky area and sensitivity of existing \hi \ surveys. However, at present a different approach is required to identify new objects in this regime.

Pavo \citep{Jones+2023b} is a close analog of Leo~P that was recently discovered using a convolutional neural network to identify low-mass, ``semi-resolved" (combined diffuse, unresolved light with point-like sources corresponding to the brightest stars) galaxies in the Legacy Surveys \citep{Dey+2019}. Its stellar mass is a few times higher than Leo~P's \citep[$\log M_\ast/\mathrm{M_\odot} = 6.1$, versus 5.4;][]{McQuinn+2024}, but it is even more isolated at a distance of $2.16^{+0.08}_{-0.07}$~Mpc (measured via the tip of the red giant branch in Hubble Space Telescope imaging; B.~Mutlu-Pakdil et al. in prep.). Although Pavo is within the Local Sheet, it is in a direction away from any Local Volume group and its nearest known neighbors are IC~4662 and IC~5152, both just over 700~kpc away.\footnote{We have updated these neighbors relative to \citet{Jones+2023b}, based on the revised distance. Previously, IC~5152 was thought to be the nearest neighbor, just over 600~kpc away from Pavo.} However, unlike Leo~P \citep{Rhode+2013,McQuinn+2015b}, Pavo has no \hii \ regions, no H$\alpha$ emission, and no stars that are clearly younger than $\sim$150~Myr in its color--magnitude diagram (CMD). 

The star formation rates (SFRs) of galaxies at these masses ($M_\ast \lesssim 10^6$~\Msol) are expected to continuously sputter as the feedback from even a couple of supernovae (SNe) could be sufficient to temporarily eject much of their gas reservoir (\citealt{Rey+2022}). The lack of very recent SF in Pavo could, in theory, provide a window where the \hi \ reservoir is largely undisturbed and its kinematics are a reliable tracer of the underlying DM potential (\citealt{Rey+2024}). 
This is of crucial importance to constrain the nature of DM. Even though stellar feedback-induced DM cores can form in such small systems if star formation proceeds unimpeded for billions years \citep[e.g.][]{Read+2016}, they are not predicted for cosmological star formation histories at this mass scale \citep[e.g.][]{Teyssier+2013, DiCintio+2014, Pontzen+2014, Bullock+2017, Lazar+2020, Orkney+2021, Sales+2022, Muni+2025}.
Thus, strong evidence for a core in the gas kinematics of a galaxy in this mass regime could be difficult to explain within the standard feedback core formation models that are used to resolve the cusp/core problem \citep[for reviews see][]{Bullock+2017,Sales+2022}. This potential opportunity motivated us to pursue Pavo's \hi \ content with follow-up observations with MeerKAT. 

In the following section we describe our MeerKAT observations of Pavo. In \S\ref{sec:results} we present our results, which are discussed further in \S\ref{sec:dicussion}. Finally, we draw our conclusions in \S\ref{sec:conclusions}. Throughout this work we adopt a distance to Pavo of 2.16~Mpc (B.~Mutlu-Pakdil et al. in prep.).

\section{Observations} \label{sec:observations}

Pavo was observed twice under the MeerKAT director's discretionary time (DDT) project DDT-20231121-MJ-01 (PI: M.~Jones). The first observations (1~h on-source) were taken during daytime in November 2023. Although \hi \ line emission from Pavo was weakly detected, the data were heavily impacted by solar interference (despite Pavo being $\sim$55$^\circ$ away from the Sun at the time) and reliable measurements of its neutral gas properties were not possible. Pavo was thus re-observed in October 2024 during nighttime, this time for 3~h on-source, when it was between approximately 40$^\circ$ and 20$^\circ$ elevation. The data were recorded simultaneously in the 4k and 32k correlator modes, the former with 856~MHz bandwidth and the latter with 107~MHz. 

The 4k data were downloaded first and reduced using the Containerized Automated Radio Astronomy Calibration \citep[\texttt{CARACal};][]{CARACal} in order to search for any detection and identify the frequency of Pavo's \hi \ line emission. The data were reduced using standard flagging, cross-calibration, self-calibration, $uv$ continuum subtraction, and imaging tasks within \texttt{CARACal}. Once the \hi \ line emission of Pavo was identified, a $\sim$5~MHz portion of the 32k data, centered on the emission frequency, was then downloaded. These data have a channel width of 3.265~kHz ($\sim$0.7~\kms), which we averaged over three channels to reduce the computational load of the data reduction. These data were then also reduced with \texttt{CARACal}. 

The final image was produced using a Briggs robust parameter of 0.5 and a 10\arcsec \ taper, giving a synthesized beam size of 20.5\arcsec$\times$9.7\arcsec, which, to improve S/N, we circularized to 21\arcsec \ (FWHM) by applying 2-dimensional Gaussian smoothing to the image. Finally, during imaging we also binned the spectral channels by a further factor of two in order to improve S/N, giving a final channel width of 19.59~kHz ($\sim$4~\kms). Although this choice limits detail of the velocity structure of the line emission, it was necessary to obtain reliable channel maps without sacrificing angular resolution and, as is discussed in the following sections, the velocity information would be limited regardless. The final image cube has an rms noise of 0.45~mJy/beam, which equates to a 3$\sigma$ column density sensitivity of $2.2\times10^{19}\;\mathrm{cm^{-2}}$ (0.18~\Msol~pc$^{-2}$) over 10~\kms.

\section{Results} \label{sec:results}

In the following subsections we describe Pavo's \hi \ properties and how these were derived from the MeerKAT observations. Pavo's \hi \ properties are also summarized in Table~\ref{tab:HIprops}.

\begin{table}[]
    \centering
    \caption{\hi \ Properties of Pavo}
    \begin{tabular}{lc}
    \hline\hline
    Parameter       &  Value\\ \hline
    RA & 19:55:01 \\
    Dec. & -61:04:31 \\
    $v$/\kms                 & $223.7 \pm 3.6$ \\
    $W_{50}$/\kms            & $19.7 \pm 2.5$ \\
    $S_\mathrm{HI}$/Jy~\kms  & $0.56 \pm 0.02$ \\
    $D_\mathrm{HI}$/kpc      & $\sim$740$^\dagger$ \\
    $\log M_\mathrm{HI}$/\Msol & $5.79 \pm 0.05$ \\
    $\log M_\mathrm{\ast}$/\Msol & $6.08^{+0.14}_{-0.04}$$^\ddagger$ \\
    $\log M_\mathrm{HI}/M_\ast$ & $-0.3 \pm 0.1$ \\
    \hline
    \end{tabular}
    \tablecomments{Coordinates indicate the center of Pavo's stellar body, not its \hi \ distribution. Absolute quantities use a distance of 2.16~Mpc.\\$\dagger$ No inclination correction has been made to $D_\mathrm{HI}$.\\$\ddagger$ This revised stellar mass estimate is based on recent HST observations of Pavo (Mutlu-Pakdil et al. in prep.).}
    \label{tab:HIprops}
\end{table}

\subsection{\hi \ morphology} \label{sec:HImorph}

\begin{figure}
    \centering
    \includegraphics[width=\linewidth]{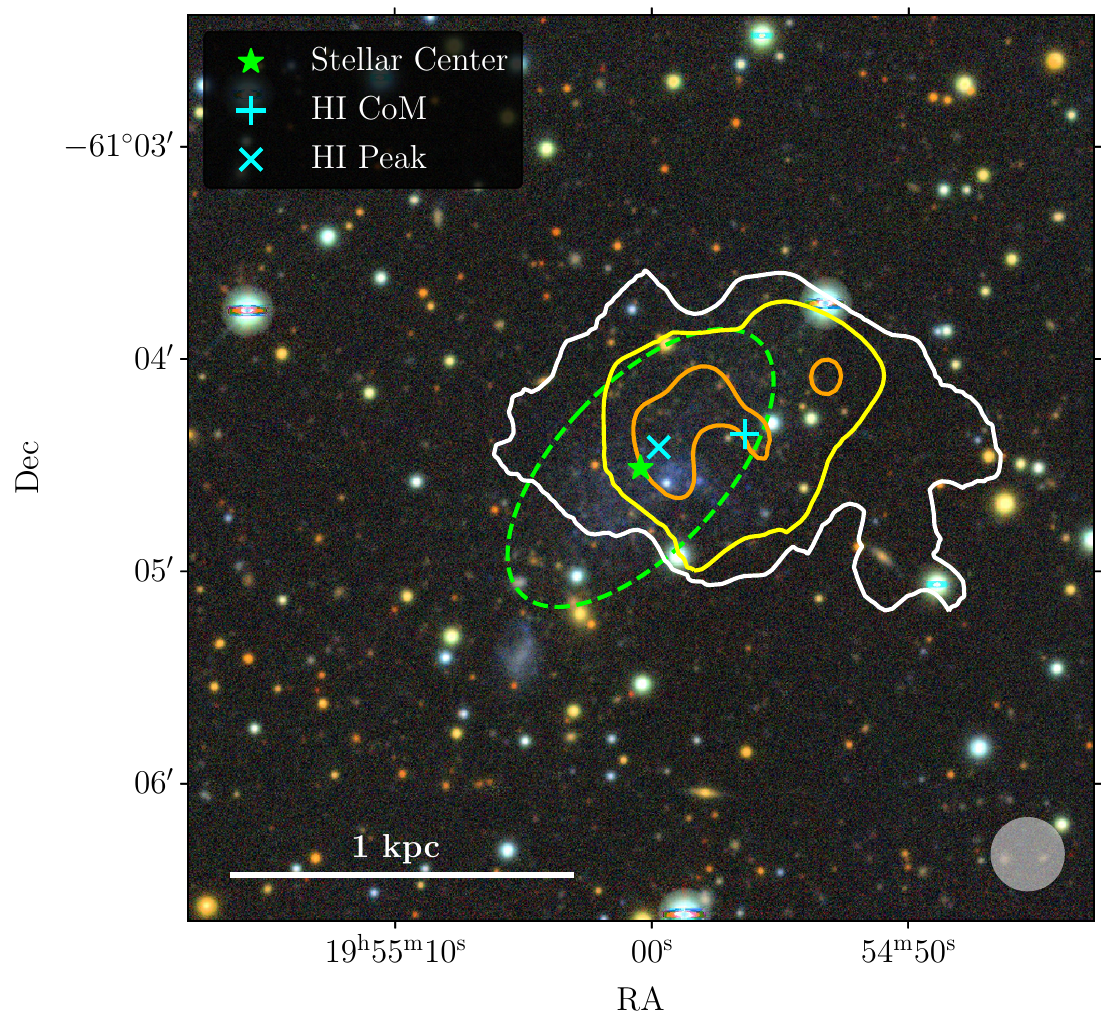}
    \caption{MeerKAT \hi \ moment zero contours overlaid on a DECaLS $griz$ image of Pavo. The green star and dashed ellipse show the center and half-light extent of Pavo's old stellar component (from HST imaging; Mutlu-Pakdil et al. in prep.), while the blue $\times$ and $+$ show the peak of the \hi \ distribution and its center of mass, respectively. Note that the bright point source near the center of Pavo is a foreground star with non-zero proper motion \citep{Jones+2023b}. The 21\arcsec \ beam is shown in the lower right corner. Contour levels begin with the white contour at 0.25~\Msol~pc$^{-2}$ ($3.1\times10^{19}\;\mathrm{cm^{-2}}$) and each subsequent contour level is double the previous.}
    \label{fig:HIoverlay}
\end{figure}

Using the smooth and clip algorithm in the Source Finding in Astronomy tool \citep[\texttt{SoFiA};][]{Serra+2015,Westmeier+2021}, we created a moment zero map of Pavo's \hi \ emission with a 5$\sigma$ threshold. The spatial smoothing scales were set to approximately one and two times the beam size, and spectral smoothing to one, two, and three ($\sim$4~\kms) channels. This map was then overlaid on a Dark Energy Camera Legacy Survey \citep[DECaLS;][]{Dey+2019} image to compare the \hi \ and optical morphology of Pavo (Figure~\ref{fig:HIoverlay}). 

We see immediately that the \hi \ distribution does not follow the underlying stellar body, but instead is concentrated on its western side, extending well beyond the stellar body in that direction, but not to the east. The distribution also appears to have an irregular and clumped structure, with multiple higher column density peaks. Although the global peak of the \hi \ moment zero map (blue ``$\times$" in Figure~\ref{fig:HIoverlay}) is almost coincident with the center of Pavo's stellar body (as determined by recent HST imaging; B.~Mutlu-Pakdil et al. in prep.), the center of mass of the \hi \ gas (blue ``$+$") is significantly offset.\footnote{The locations of the \hi \ center of mass and peak are calculated in \S\ref{sec:HIoffset} and are discussed further in that section.} 

\begin{figure*}
    \centering
    \includegraphics[width=\linewidth]{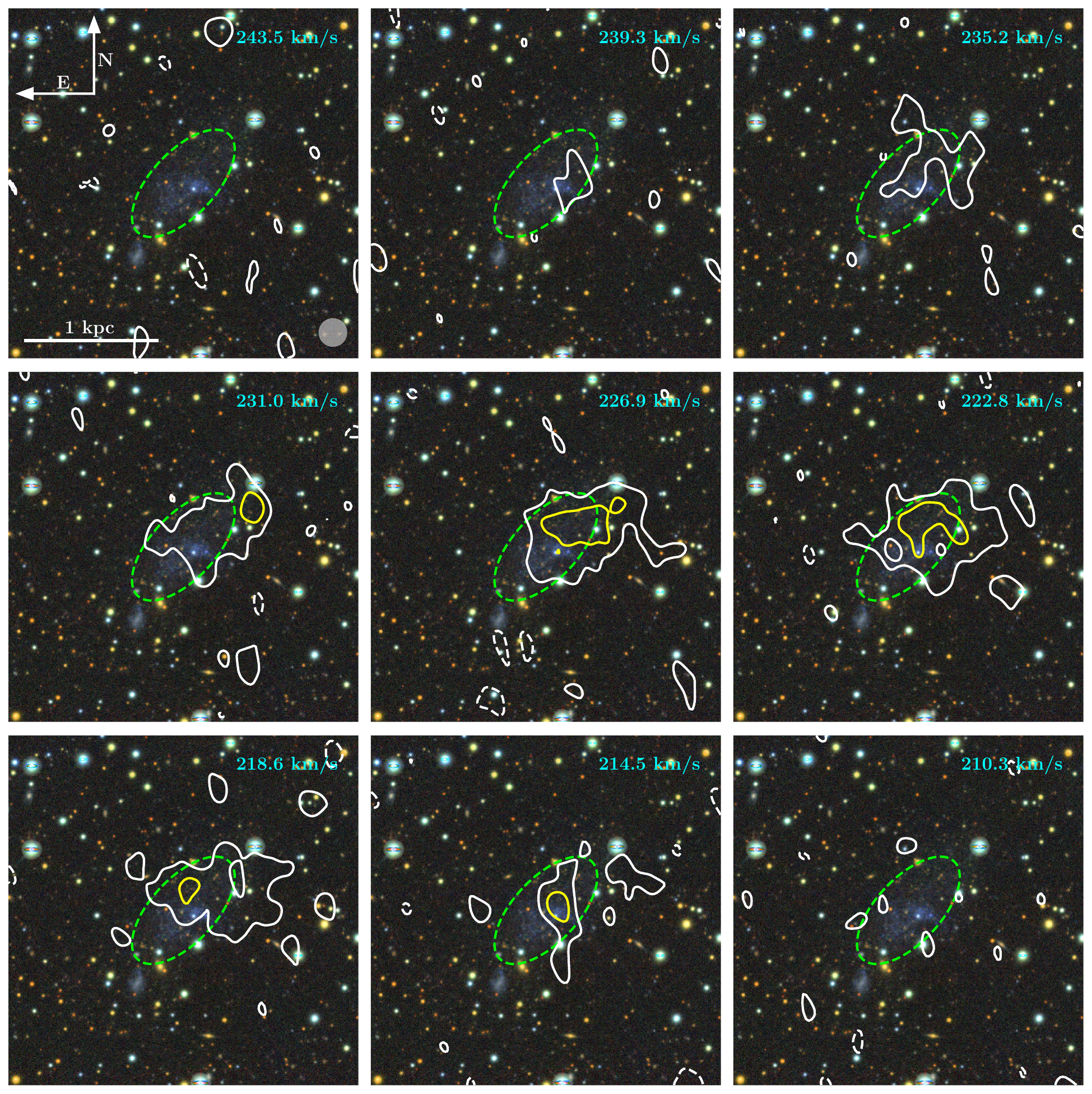}
    \caption{\hi \ channel maps of Pavo overlaid on a DECaLS $griz$ image. The green dashed ellipse represents the stellar body (as in Figure~\ref{fig:HIoverlay}). The contour levels are -2.5$\sigma$, 2.5$\sigma$, and 5$\sigma$ in each channel, where 2.5$\sigma$ is $1.2\times10^{-19}\;\mathrm{cm^{-2}}$ (over 4~\kms).}
    \label{fig:chan_maps}
\end{figure*}

We attempted to identify clear signs of rotation in Pavo with both a moment one map and a position--velocity (PV) slice. The moment one map, even with a higher S/N threshold, is too noisy to see any meaningful structure. The PV slice is also noisy and it is difficult to identify a clear gradient. 
The \hi \ channel maps (Figure~\ref{fig:chan_maps}) show the clearest evidence for a velocity gradient that is roughly aligned with the major axis of the stellar body, with the gas at $\sim$230~\kms \ being concentrated on the NW of the galaxy and retreating towards its (stellar) center at lower velocities ($\sim$215~\kms). However, this is truncated as the \hi \ distribution does not extend out to the SE side of the galaxy (Figure~\ref{fig:HIoverlay}).

\subsection{\hi \ mass and velocity width} \label{sec:HImass}

\begin{figure}
    \centering
    \includegraphics[width=\linewidth]{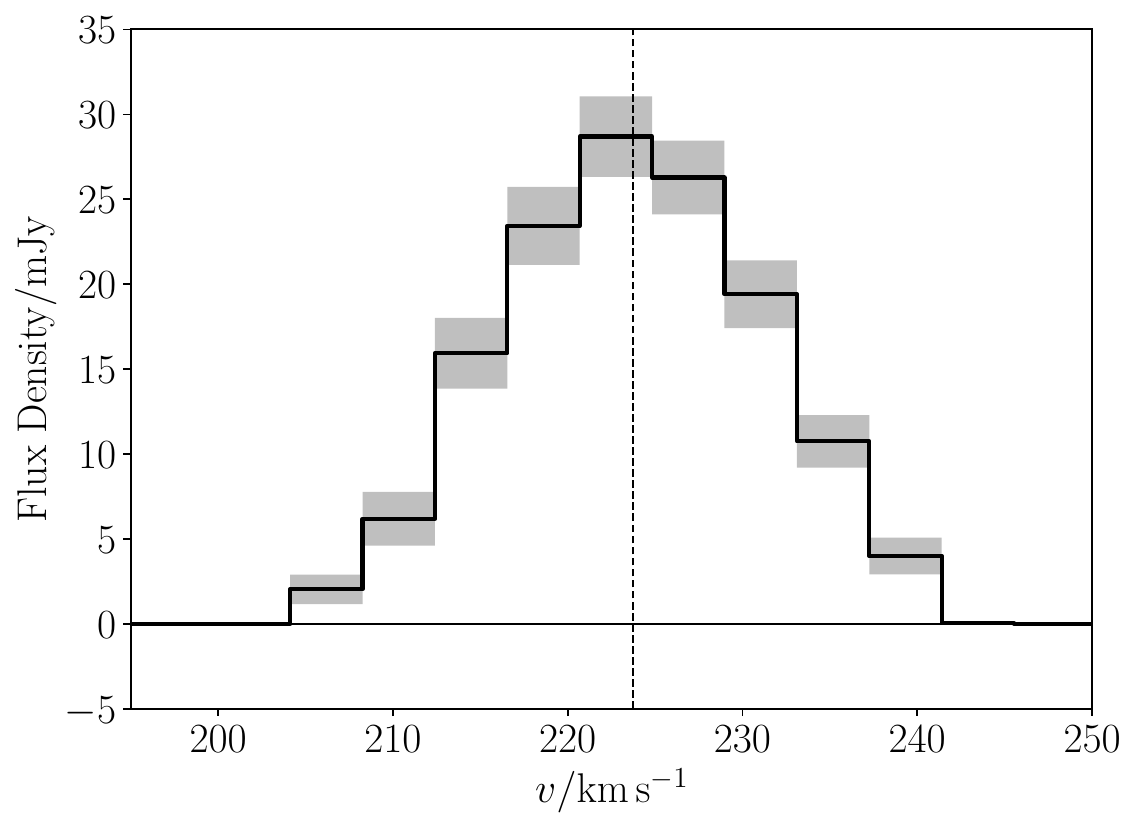}
    \caption{Spectral profile of Pavo's \hi \ line emission. The dashed vertical line shows the flux-weighted line center (at 224~\kms). The grey shading indicates the flux density uncertainty in each channel. The velocity axis follows the optical convention (i.e. $v=cz$) and is in the barycentric frame.}
    \label{fig:HIspec}
\end{figure}

The \hi \ emission within the mask discussed in \S\ref{sec:HImorph} was summed over the two spatial dimensions to produce the global line profile (Figure~\ref{fig:HIspec}). This profile is roughly Gaussian in shape, as is often the case for low-mass galaxies. The expected circular velocities of the low-mass halos that host galaxies such as Pavo are only on the order 10-20~\kms \ \citep{Rey+2024}. Thus, the observed line width, $W_{50} = 19.7 \pm 2.5$~\kms, could be consistent with a combination of rotation and turbulence, or it could be dominated by random motions, given its disturbed morphology. The flux-weighted line center is $223.7 \pm 3.6$~\kms \ (barycentric), the first measurement of Pavo's radial velocity.

Collapsing the final dimension of the line profile gives the global \hi \ flux of Pavo as $0.56 \pm 0.02$~Jy~\kms. Following the standard conversion \citep[e.g.][]{Giovanelli+2005}, this equates to a total \hi \ mass of $\log M_\mathrm{HI}/\mathrm{M_\odot} = 5.79 \pm 0.05$ at 2.16~Mpc. The uncertainty on the \hi \ mass includes the statistical uncertainties for the flux and distance, plus an assumed 10\% uncertainty in the absolute flux calibration. This makes Pavo, to date, the lowest \hi \ mass galaxy known in isolation and with a robust distance measurement. Pavo is also significantly more \hi-poor than other related objects, with a neutral gas fraction of $\log M_\mathrm{HI}/M_\ast = -0.3$ compared to 0.3, 0.2, and 0.6 for Leo~T, Leo~P, and Corvus~A, respectively \citep{Weisz+2012,McQuinn+2015b,Adams+2018,Jones+2024}. The exception is Antlia~B, for which $\log M_\mathrm{HI}/M_\ast \approx -0.35$ \citep{Sand+2015b}, however, this is in close proximity to NGC~3109 and may be interacting with it.

\subsection{\hi \ size--mass relation} \label{sec:size-mass}

\begin{figure}
    \centering
    \includegraphics[width=\linewidth]{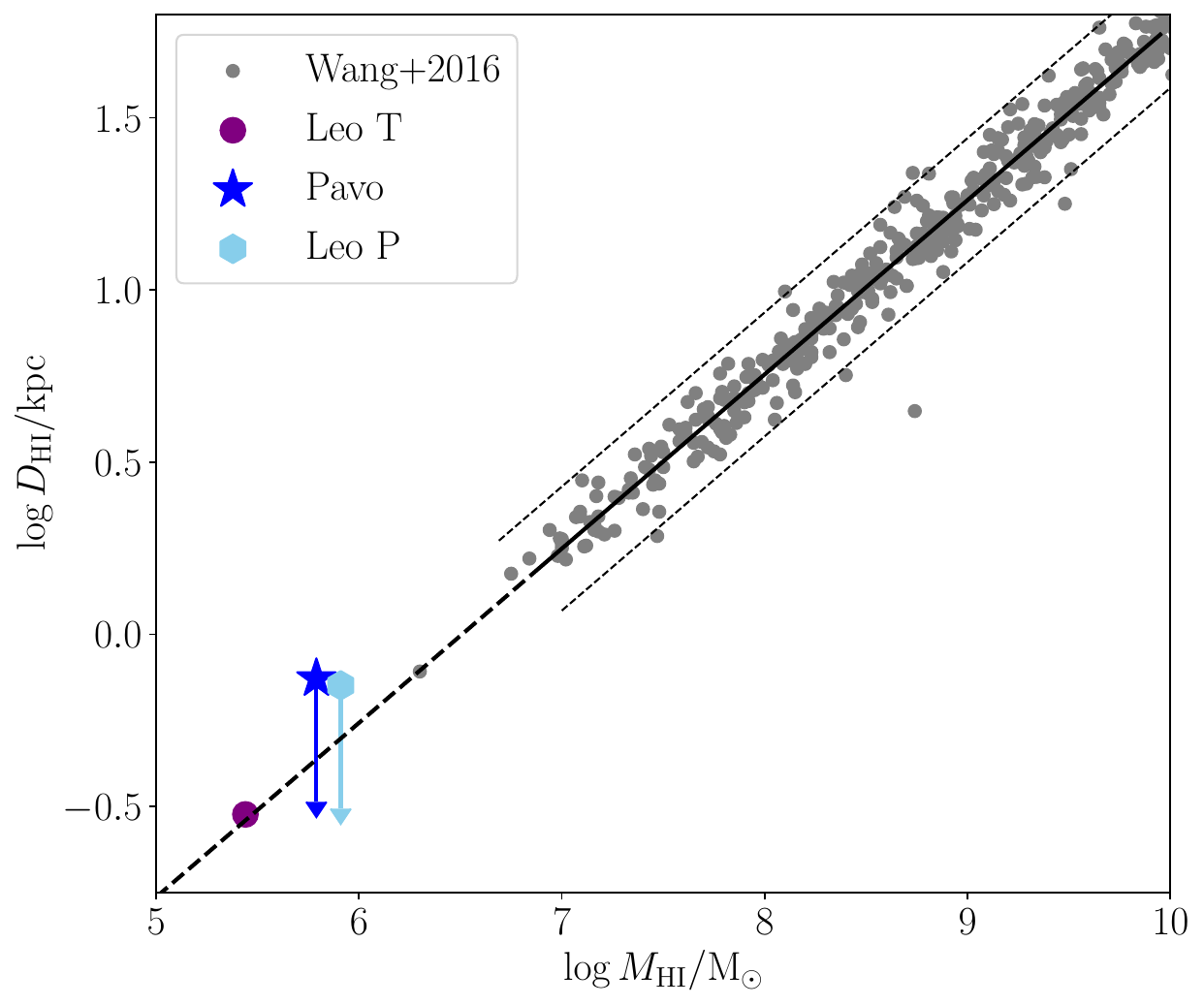}
    \caption{Pavo (blue star), Leo~P (light blue hexagon), and Leo~T (purple circle) on the \hi \ size--mass relation (black solid line) of \citet{Wang+2016}, with the thin dashed black lines showing the 3$\sigma$ scatter about the relation. The \hi \ diameter values for Pavo and Leo~P are plotted at their observed values without correction for inclination. The length of the arrows indicate the approximate maximum shifts that the inclination corrections could apply. The grey data points are the galaxies compiled and uniformly re-analyzed by \citet{Wang+2016}, originally from: \citet{Verheijen+2001,Swaters+2002,Noordermeer+2005,Begum+2008,Koribalski+2008,Walter+2008,Chung+2009,Kovac+2009,Hunter+2012,Serra+2012,Serra+2014,Wang+2013,Lelli+2014,Martinsson+2016}.}
    \label{fig:size-mass}
\end{figure}

The \hi \ size--mass relation \citep[e.g.][]{Broeils+1997,Swaters+2002,Wang+2016} is a tight scaling relation between a galaxy's \hi \ mass and the diameter of its \hi \ distribution. It has been shown to hold over more than four orders of magnitude in \hi\ mass \citep{Wang+2016}, from low-mass dwarf irregulars to Milky Way-mass spirals. 
From a sample of more than 500 nearby galaxies with \hi \ interferometric observations, \citet{Wang+2016} found only seven strong outliers from this relation, all highly disturbed objects. Indeed, there is some evidence that interactions can drive galaxies slightly away from the mean relation \citep{Wang+2024}. However, many galaxies that are clearly disturbed nevertheless fall within the scatter of the relation, including the LMC and SMC. In addition, the relation seems to be obeyed by some other types of anomalous galaxies, for example, those truncated by ram pressure stripping \citep{Wang+2016} and those with abnormally large \hi \ reservoirs for their stellar mass \citep{Lutz+2018}. 
\citet{Stevens+2019} argues that the \hi \ size--mass relation is an inevitability resulting from the tendency for \hi \ radial distributions to be similar in form across all types of galaxies and the fact that \hi \ column densities effectively saturate due to the \hi--H$_2$ transition. They conclude that only galaxies hosting \hi \ reservoirs that are almost completely disrupted should be expected to strongly deviate from the \hi \ size--mass relation.

Given Pavo's irregular \hi \ morphology, it is nontrivial to measure its \hi \ size. Typically, the diameter of a galaxy's \hi \ content is measured at a column density of 1~$\mathrm{M_\odot\,pc^{-2}}$ ($1.25\times10^{20} \; \mathrm{cm^{-2}}$) when placing it on the \hi \ size--mass relation. However, at this column density, Pavo's \hi \ distribution is already lopsided and fragmented. We therefore cannot adopt the approach of \citet{Wang+2016} where an azimuthally averaged \hi \ profile is measured extending from the optical center of the galaxy. Instead we follow a similar approach to that which \citet{Adams+2018} used for Leo~T, based on image moments \citep{Banks+1995} of the \hi \ distribution. Using this approach we fit an ellipse to the \hi \ gas distribution at $\Sigma_\mathrm{HI} = 1 \; \mathrm{M_\odot\,pc^{-2}}$, equally weighting each pixel in the moment zero map above this threshold. This gives an \hi \ diameter of 74\arcsec, which we correct for broadening due to the beam as in \citet{Wang+2016}, obtaining a final \hi \ diameter for Pavo of 740~pc (71\arcsec). 

As \hi \ emission is optically thin, the major axes of inclined distributions appear larger at fixed column density than they would if face-on. For a perfect thin disk this results in a correction factor of $\cos i$ to the major axis diameter of the observed \hi \ distribution. The nature of Pavo's \hi \ morphology makes it impossible to determine a meaningful inclination estimate from the gas, and we must therefore rely on the apparent inclination of the stellar body, which we estimate to be 64$^\circ$, based on an axial ratio of 0.48 (Mutlu-Pakdil et al in prep.) and an assumed intrinsic axial ratio of 0.2 when edge-on. This is probably an overestimate of the true inclination as the stars are unlikely to be arranged in a thin disk. However, as described below, this inclination estimate will be used only to determine an upper limit on the correction factor for Pavo's \hi \ diameter. Thus, a slight overestimate is tolerable in this regard.

In Figure~\ref{fig:size-mass} we show the \hi \ size--mass relation of \citet{Wang+2016} with Leo~T and Pavo highlighted. For comparison we also calculated the \hi \ diameter of Leo~P (710~pc or 91\arcsec) based on the VLA \hi \ observations of \citet{Berstein-Cooper+2014}, following the same process as that described above for Pavo. The stellar body of Leo~P also has an identical axial ratio \citep{McQuinn+2015b} to Pavo, and thus an equivalent inclination correction. The points for Pavo and Leo~P are plotted at their observed values (after correcting for beam smearing), but no correction for inclination, thus representing upper limits on their \hi \ diameters. The length of the downward arrow from each point indicates the maximum expected shift that an inclination correction could impose. In practice, it is unclear if the gas is at similar inclination to the stellar body and it is also implausible that Pavo's gas is arranged in a thin disk. Both are likely to reduce the level of the inclination correction, hence why we suggest that the arrows are viewed as the maximum plausible shifts.

The observed values for both Pavo and Leo~P are on the upper extreme of the 3$\sigma$ scatter region for the \citet{Wang+2016} \hi \ size--mass relation. However, given the uncertainty in the inclination corrections, it is still extremely likely that, if face-on, both Pavo and Leo~P would be entirely consistent with this relation. This is not to imply that the gas distribution of Pavo is not disturbed, but rather agrees with the conclusion of \citet{Stevens+2019}, that the \hi \ reservoir of a galaxy must be almost entirely disrupted in order for it to strongly deviate from the \hi \ size--mass relation. In addition, the locations of Leo~T, Pavo, and Leo~P suggest that this relation likely continues unbroken down to the lowest mass star-forming galaxies \citep[unlike the stellar size--mass relation, e.g.,][]{Chamba+2024}. However, additional high-resolution \hi \ observations of galaxies in this mass regime are needed to confirm this.

\subsection{Offset of \hi \ distribution from stars} \label{sec:HIoffset}

\citet{Rey+2022} showed that galaxies with similar stellar masses to Pavo in the Engineering Dwarfs at Galaxy formation's Edge \citep[\textsc{edge};][]{Agertz+2020, Rey+2025} simulations frequently exhibit large offsets (up to $\sim$500~pc) between their stellar and \hi \ centers as stellar feedback disturbs the gas reservoir. Pavo's complex \hi \ distribution makes determining a meaningful center challenging and we therefore adopt two approaches. If we simply take the global maximum of the \hi \ column density as the \hi \ center then we find a relatively minor offset from the center of the stellar body (Mutlu-Pakdil et al. in prep.) of 82~pc (in projection). This value is comfortably within the typical range for \textsc{edge} dwarfs \citep[][figure~3]{Rey+2022}. If instead we take the intensity-weighted center of the moment zero map as the \hi \ center then we find a much larger projected offset of 320~pc. This value is also within the distribution of values reported in \textsc{edge}, although on the higher end and would suggest that Pavo might be in a particularly tumultuous period in its life-cycle. We conclude that the observed offset is quantitatively compatible with predictions from simulations of low-mass dwarfs.

\section{Discussion} \label{sec:dicussion}

\begin{figure*}
    \centering
    \includegraphics[width=\textwidth]{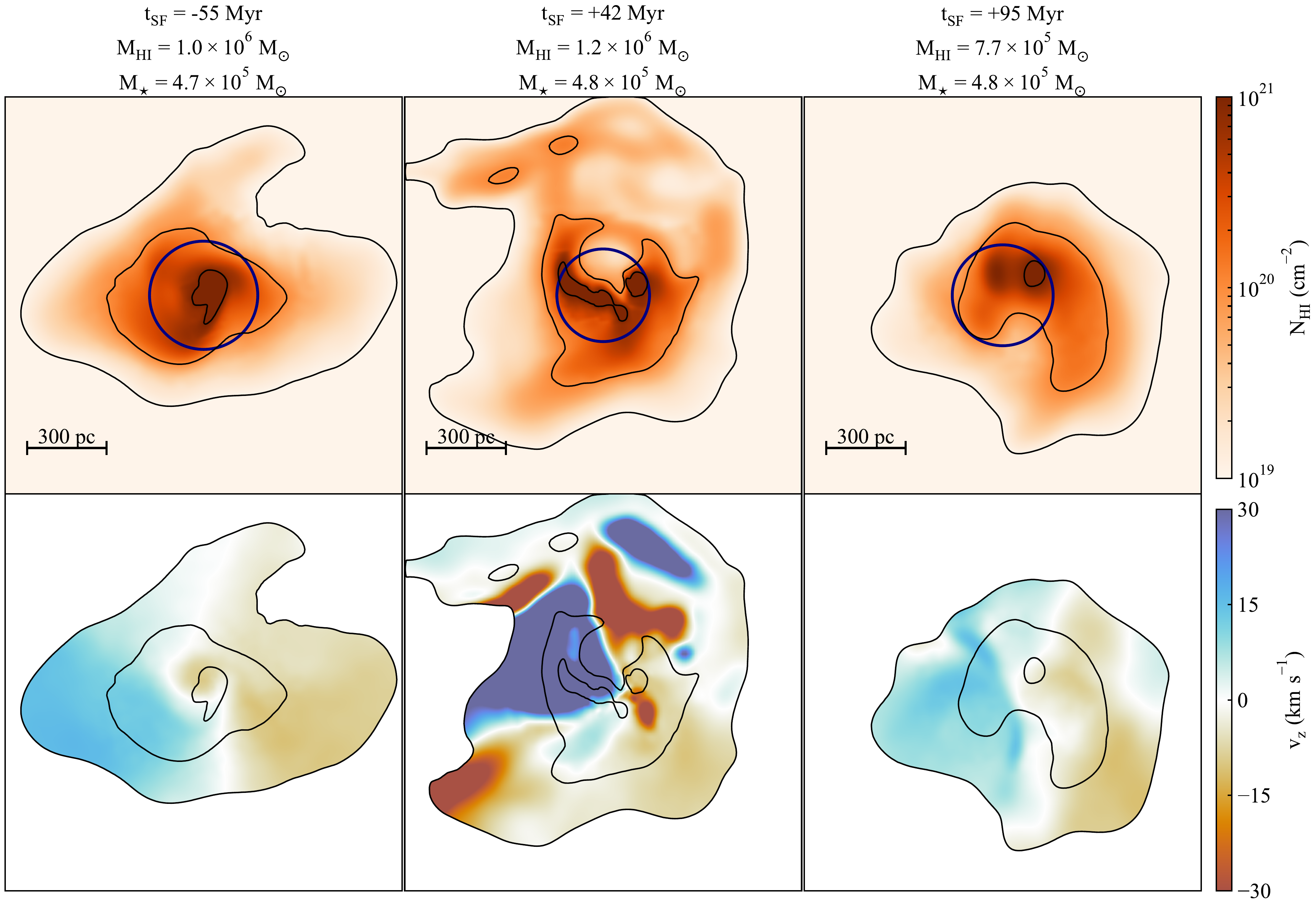}
    \caption{Time sequence of \hi \ maps (top) and line-of-sight gas velocity fields (bottom) from an \textsc{edge} simulated low-mass dwarf similar to Pavo (at an inclination of 60$^\circ$). Before a SF event (left), the galaxy's \hi \ content is centrally concentrated, aligned with the stellar body (blue circle showing the V-band projected half-light radius), and shows relatively ordered rotation ($\approx$10~\kms). Shortly after SF (middle panel), strong outflows disrupt the gas component, creating holes in the \hi \ distribution. After $\approx$100~Myr (roughly the timing of Pavo; right panel), the \hi \ reservoir remains offset from the stellar body and gaps carved by the previous outflows remain visible. At later times inflows become more prevalent again, eventually rebuilding the \hi \ reservoir and restarting the cycle before the next SF event. This scenario is generic across the \textsc{edge} simulations (see \citealt{Rey+2022, Rey+2024} for other examples and discussion).
    }
    \label{fig:edgedwarfs}
\end{figure*}

\subsection{\hi \ morphology of Pavo compared to simulated low-mass dwarfs}

Observations of samples of low-mass, star-forming galaxies that are typically 1-2~dex more massive than Pavo or Leo~P have repeatedly shown evidence for bursty SF and disturbed \hi \ morphologies \citep[e.g.,][]{Mas-Hesse+1999,Begum+2006,Begum+2008,Cannon+2011,Hunter+2012,McQuinn+2010,McQuinn+2018,McNichols+2016}. Models and simulations have long pointed to SN feedback as the likely cause of these quasi-episodic SFRs and irregular gas distributions, with SN-driven winds removing gas from the centers of their shallow potential wells and triggering pauses in SF \citep[e.g.,][]{Dekel+1986,Spaans+1997,Stinson+2007,DiCintio+2014,Read+2016,Sands+2024}. Galaxies such as Pavo and Leo~P push these models even further, into a regime where it is plausible that feedback from just one or two SN could temporarily erase the entire \hi \ reservoir \citep{Rey+2022}. Such low mass galaxies therefore offer a unique opportunity to dissect the feedback cycle between SF and galactic outflows. 

High-resolution simulations of faint, star-forming dwarf galaxies predict that \hi \ properties will vary on short timescales ($\approx$50-100~Myr), with disturbed \hi \ morphologies, frequent offsets between the stellar and \hi \ bodies \citep{Rey+2022}, and complex kinematics cycling between inflows, outflows and short-lived settled disks \citep{Downing+2023, Rey+2024}. In such small objects, the dynamical time, disk-crossing time, and the lifetimes of massive stars exploding as core collapse SN are comparable ($\approx$20~Myr). Once SF is triggered and SN feedback occurs, galactic outflows can thus rapidly disrupt a significant fraction of the \hi \ structure, leading to disturbed morphologies and disorganized kinematics. But the short dynamical times also mean that inflows can allow for organized, smooth and regular \hi \ reservoirs to appear again after several tens of Myr. 

Figure~\ref{fig:edgedwarfs} exemplifies this cycle in the \textsc{edge} simulations, showing three \hi \ column density maps (upper panels) 55~Myr before and 42 and 95~Myr after a SF episode. Contours show \hi \ column densities of $10^{19}$, $10^{20}$ and $10^{21} \, \mathrm{cm^{-2}}$ at a resolution of approximately 7~pc. Note that this is substantially finer spatial resolution than the MeerKAT observations of Pavo, and results in peak column densities that are an order of magnitude higher than those in Figure~\ref{fig:HIoverlay}. The lower panels of Figure~\ref{fig:edgedwarfs} show the line-of-sight velocity field of the total gas (\hi \ and \hii), weighted by the \hi \ mass along each line-of-sight. A complete description of the simulation setup, the sample of simulated \textsc{edge} dwarf galaxies and their \hi \ content and kinematics can be found in \citet{Rey+2022, Rey+2024}. The dwarf galaxy shown (`Halo 600') has a DM halo mass of $M_{200} = 3.3\times10^{9}$~\Msol, and is representative of the \textsc{edge} sample of dwarf galaxies at this stellar mass (\citealt{Rey+2022, Rey+2024, Rey+2025}). We have inclined this system at 60$^\circ$ to roughly match our estimate of Pavo's inclination (\S\ref{sec:size-mass}). Simulated data is only saved every $\approx$100~Myr, so the specific time is chosen to maximize coverage around a SF event. 

None of the panels exhibit the well-ordered structure common in higher mass galaxies \citep[e.g.][]{Walter+2008}. Instead, all show \hi \ distributions (top panels) that are to some degree lopsided and offset from the projected V-band half-light radius (blue circles\footnote{The V-band half-light radius varies from panel to panel, as young stars temporarily dominate the light when SF occurs.}). However, shortly prior to the SF episode (left panel) the \hi \ is mostly centrally concentrated in a roughly disk-like configuration with clear signs of ordered rotation (bottom panels). Soon after SF (middle column) strong outflows are visible driving out gas and creating holes in the \hi \ distribution. Later this gas falls back towards the center of the potential well (and the stellar body) and reforms a disk-like configuration, closing the cycle. 

A key prediction of these simulations is that the timing of the last SF episode and SN explosions should correlate with \hi \ properties. This is qualitatively supported by contrasting Pavo and Leo~P. Although both are clearly detected in the near UV, which is sensitive to SF in the past $\sim$200~Myr \citep[e.g.][]{Lee+2009}, Leo~P has a single \hii \ region and accompanying H$\alpha$ emission \citep{Rhode+2013}, while Pavo has neither \citep{Jones+2023b}. Thus, Leo~P has been forming stars in the last 10~Myr, and it is likely at a stage roughly in between the first two panels of Figure~\ref{fig:edgedwarfs}, with SN feedback yet to fully disrupt the \hi \ reservoir. As a result, Leo~P's \hi \ is quite regular on large scales, is aligned with the newborn stars and shows tentative signs of rotation \citep{Berstein-Cooper+2014}. Corvus~A is likely in a similar situation to Leo~P. Its gas reservoir also appears to be relatively undisturbed and has tentative signs of rotation, albeit at relatively poor spatial resolution \citep{Jones+2024}. Although it does not have any H$\alpha$ emission, the youngest stars identifiable in its CMD are only $\sim$50~Myr old, indicating very recent SF. By contrast, the youngest stars in Pavo's CMD are consistent with having formed as much as 150~Myr ago. Conversely, Pavo's \hi \ reservoir is lopsided, offset from its stellar body, with a much lower gas fraction and showing a weak and truncated velocity gradient (Figure~\ref{fig:chan_maps}), all of which are reminiscent of the right panel of Figure~\ref{fig:edgedwarfs}. 

To extend this comparison further, we can also consider the slightly lower mass Leo~T.
Unlike both Pavo and Leo~P, Leo~T is almost undetectable \citep{Lee+2011} in NUV (despite being significantly nearer than either), indicating that it has likely not formed many stars in the past $\sim$200~Myr.\footnote{Although \citet{Weisz+2012} fit a star formation history for Leo~T that does not shut off until $\sim$25~Myr ago, the extremely low mass of Leo~T means that the high mass end of the stellar initial mass function is very poorly sampled, introducing a considerable stochastic uncertainty to recent SFR estimates. That said, the youngest stars visible in the CMD in \citet{Weisz+2012} appear to be consistent with stellar isochrones as old as 500~Myr. Similarly, the relative lack of UV emission also points to very little SF in the past few hundred Myr.} Leo~T's \hi \ reservoir also appears to be regular in shape and kinematically quiet \citep[it is uncertain if there is rotation as the \hi \ disk is almost face-on;][]{Adams+2018}. Thus, we suggest that Leo~T has come full circle and is most akin to the first panel of Figure~\ref{fig:edgedwarfs}, where gas has re-settled into a disk long after the most recent SF episode. Thus, Leo~P, Pavo, and Leo~T appear to represent three distinct phases of this process.

If Pavo is in the intermediate stage between a SF episode and \hi \ re-settling, then its disturbed gas reservoir is likely a sign that its gas is either in the process of being blown out by feedback or falling back toward the stellar body. With the available data, it is difficult to conclusively tell whether the disturbances reflect outflows from the last SF episode or inflows starting to rebuild a smooth \hi \ reservoir. There is a weak, but visible, velocity gradient from the NW side of Pavo to its center (Appendix~\ref{sec:chan_maps}) suggestive of rotation. This could point to Pavo's morphology being just half of an \hi \ disk, with the other half having been disrupted by feedback. This aligns with the fact that the recent SF in Pavo is near the center and towards the SE. The SE side of a potential \hi \ disk could have been ionized and expelled by feedback, with the NW side yet to be disturbed.
However, the kinematic gradient across the NW does not continue to the other side, since no \hi \ is detected there, making it difficult to be fully conclusive about the existence of a disk. Furthermore, we suspect that $\sim$150~Myr has elapsed since the last SF episode, more than a couple of orbital dynamical times. SN-outflows can locally evacuate the \hi \ at high column densities (e.g. Figure~\ref{fig:edgedwarfs}, middle panel), but such features are expected to smooth out at low column densities after $\approx$100~Myr. Thus, explaining Pavo's gas morphology and kinematics as being half a disk would imply an improbable timing, long enough after the most recent SF that part of the disk could have been disrupted, but not long enough that orbital motions have not smoothed out the disturbance. We therefore favor an interpretation where we are seeing Pavo somewhat later after the most recent SF (more compatible with the age of the youngest stars visible in the CMD) and the gas is falling back towards the center of the potential well.

\subsection{Toward gas kinematics in extremely low mass dwarfs}

Given the present state of Pavo's \hi \ reservoir discussed above, it is unlikely that even with deeper and higher resolution follow-up (for example, with the Square Kilometre Array) meaningful dynamical information about its DM halo could be extracted from its gas kinematics \citep[e.g.][]{Downing+2023}. Although Pavo appears not to be a suitable target, high resolution gas kinematics of even a single galaxy in this mass regime with relatively well-ordered rotation would represent a powerful test for current $\Lambda$CDM-based models of low-mass galaxy formation. The lifetime stellar feedback in galaxies of Pavo's mass should not impart sufficient energy to the DM halo to force it into a cored configuration \citep[e.g.][]{Teyssier+2013, DiCintio+2014, Lazar+2020, Orkney+2021, Muni+2025}. Thus, if the \hi \ rotation curve of such a galaxy reflected a cuspy DM profile, this would be a powerful verification for this paradigm. 

With this goal in mind, we suggest that the best path forward is to identify isolated, extremely low mass ($M_\ast \lesssim 10^6$~\Msol) galaxies that have not undergone significant SF in the past few hundred Myr (see discussion in \citealt{Rey+2024}). We are already aiming to find such galaxies through the SEmi-Automated Machine LEarning Search for Semi-resolved galaxies (SEAMLESS; \citealt{Jones+2023b,Jones+2024}, Fielder et al in prep.). \citet{Giovanelli+2015} argued that these quiescent field dwarfs should be numerous, however, their abundance is directly related to the duty cycle of SF episodes and feedback. \citet{Rey+2024} find that although ordered rotation can dominate the gas kinematics in systems analogous to Pavo, these phases are short-lived (their figure 3). In practice, such objects may prove to be too rare for one to be found using existing surveys. Fortunately, we anticipate that Euclid, Roman, and Rubin will greatly expand the available discovery space in the near future with both resolved and semi-resolved searches \citep[e.g.][]{Mutlu-Pakdil+2021}. 

Another avenue for discovery of these objects may be ongoing wide-area \hi \ surveys. If Pavo had been within the footprint of the ALFALFA survey it would have been detected at high S/N, despite its low \hi \ mass. The Apertif imaging survey \citep{Adams+2022} recently completed \hi \ observations of $\sim$1000~sq~deg of the sky, including extensions north of the ALFALFA footprint, the Widefield ASKAP L-band All-sky Blind SurveY \citep[WALLABY;][]{Koribalski+2020} project is currently mapping the entire southern sky at a similar depth to ALFALFA, and the ongoing FAST All Sky \hi \ (FASHI) survey \citep{Zhang+2024} will exceed both the depth and coverage of ALFALFA in the northern hemisphere.

\section{Conclusions} \label{sec:conclusions}

We detect the \hi \ reservoir of the recently discovered low-mass galaxy Pavo with MeerKAT DDT observations. The integrated flux of its \hi \ line emission is $0.56 \pm 0.02$~Jy~\kms, which at its distance of $2.16^{+0.08}_{-0.07}$~Mpc (Mutlu-Pakdil et al. in prep.) equates to a total \hi \ mass of $\log M_\mathrm{HI}/\mathrm{M_\odot} = 5.79 \pm 0.05$. This makes Pavo's \hi \ reservoir the lowest mass ever detected in any isolated galaxy beyond the LG (with a robust distance measurement). 

Although Pavo is remarkably isolated, with no known galaxy within over 700~kpc, we find that its gas morphology is highly disturbed. The \hi \ center of mass is offset from its stars, there are multiple peaks in the \hi \ distribution, and lopsided tails out of one side of the galaxy. There are weak signs of a continuous velocity gradient over most of the gas reservoir, but the velocity field is too disorganized for a rotation curve analysis.

The properties of Pavo's \hi \ reservoir are remarkably consistent with those of analogous objects in the \textsc{edge} simulations \citep{Rey+2020,Rey+2022, Rey+2024}. We posit that the nearby, low-mass, gas-bearing galaxies Leo~P, Pavo, and Leo~T represent three distinct stages of the boom-and-bust SF cycle of low-mass galaxies. Leo~P with its massive and centrally concentrated \hi \ reservoir, coupled with H$\alpha$ emission and an \hii \ region, is currently experiencing a SF episode that has not yet had sufficient time to disrupt its gas reservoir. Whereas we appear to be viewing Pavo 100 Myr or so after the last SF episode, with no H$\alpha$ emission or extremely young stars in its CMD. Feedback has thus had time to disrupt and ionize much of its gas, resulting in a feeble and highly disturbed \hi \ reservoir. Given the age of the youngest stars visible in Pavo's CMD, $\sim$150~Myr (\citealt{Jones+2023b}, Mutlu-Pakdil et al. in prep.), we suspect that we are currently witnessing Pavo as its gas is in the process of falling back towards its stellar body. Finally, it has likely been even longer since Leo~T's most recent SF episode and its gas has had sufficient time to fall back to the center of its potential well and re-settle into a regular disk configuration.

Given these findings, and with the long-term goal of using \hi \ kinematics to directly probe the inner structure of the DM halos of extremely low mass galaxies, we suggest that efforts should be focused on identifying isolated galaxies in this mass range (i.e. $M_\ast \lesssim 10^6$~\Msol) that have not formed stars for $\gtrsim$200~Myr. These are most likely to host relatively well-ordered gas reservoirs where \hi \ kinematics will accurately reflect the underlying DM halo. Such galaxies may be identifiable in existing wide-field ground-based surveys, but we would likely need an element of luck on our side, as they are also likely to be quite rare \citep[because the ordered disk phase is expected to be short-lived;][]{Rey+2024}. However, if such galaxies exist, they will certainly be discoverable in the latest ongoing and upcoming optical/IR surveys (Euclid, Roman, and Rubin), as well as ongoing \hi \ wide-field surveys (WALLABY and FASHI).

\begin{acknowledgments}
We thank John Cannon for providing the VLA \hi \ data cube of Leo~P from \citet{Berstein-Cooper+2014}.
The MeerKAT telescope is operated by the South African Radio Astronomy Observatory, which is a facility of the National Research Foundation, an agency of the Department of Science and Innovation.
The data published here have been reduced using the CARACal pipeline, partially supported by ERC Starting grant number 679627 ``FORNAX'', MAECI Grant Number ZA18GR02, DST-NRF Grant Number 113121 as part of the ISARP Joint Research Scheme, and BMBF project 05A17PC2 for D-MeerKAT. Information about CARACal can be obtained online under the URL: \url{https://caracal.readthedocs.io}.
This work used images from the Dark Energy Camera Legacy Survey (DECaLS; Proposal ID 2014B-0404; PIs: David Schlegel and Arjun Dey). Full acknowledgment at \url{https://www.legacysurvey.org/acknowledgment/}. 
DJS acknowledges support from NSF grant AST-2205863.
KS acknowledges support from the Natural Sciences and Engineering Research Council of Canada (NSERC).
BMP acknowledges support for the program HST-GO-17514 provided by NASA through a grant from the Space Telescope Science Institute, which is operated by the Association of Universities for Research in Astronomy, Inc., under NASA contract NAS5-26555.
AK acknowledges support from NSERC, the University of Toronto Arts \& Science Postdoctoral Fellowship program, and the Dunlap Institute.
DZ acknowledges support from NSF AST-2006785 and NASA ADAP 80NSSC23K0471.
\end{acknowledgments}

%

\vspace{5mm}
\facilities{MeerKAT, Blanco, VLA}


\software{\href{https://caracal.readthedocs.io/}{\texttt{CARACal}} \citep{CARACal}, \href{https://gitlab.com/SoFiA-Admin/SoFiA-2}{\texttt{SoFiA~2}} \citep{SoFiA2},
\href{https://www.astropy.org/index.html}{\texttt{astropy}} \citep{astropy2013,astropy2018}, \href{https://photutils.readthedocs.io/en/stable/}{\texttt{Photutils}} \citep{photutils}, \href{https://reproject.readthedocs.io/en/stable/}{\texttt{reproject}} \citep{reproject}, \href{https://matplotlib.org/}{\texttt{matplotlib}} \citep{matplotlib}, \href{https://numpy.org/}{\texttt{numpy}} \citep{numpy}, \href{https://scipy.org/}{\texttt{scipy}} \citep{scipy1,scipy2}, \href{https://pandas.pydata.org/}{\texttt{pandas}} \citep{pandas2},  \href{https://sites.google.com/cfa.harvard.edu/saoimageds9}{\texttt{DS9}} \citep{DS9}, \href{https://github.com/pynbody/genetIC}{\texttt{genetIC}} \citep{Stopyra+2021}, \href{https://github.com/ramses-organisation/ramses}{\texttt{RAMSES-RT}} \citep{Teyssier2002, Rosdahl+2013}, \href{https://github.com/pynbody/pynbody}{\texttt{pyNbody}} \citep{Pontzen+2013}, \href{https://github.com/pynbody/tangos}{\texttt{Tangos}}, \citep{Pontzen+2018}}




\bibliography{refs}{}
\bibliographystyle{aasjournal}



\end{document}